\documentclass[aps,nofootinbib,showpacs,showkeys,preprint
] {revtex4}

\usepackage{epsf,epsfig,subfigure,axodraw,graphicx,amsmath,amssymb}
\usepackage{color}

\begin{document}
 \begin{flushright}
{\tt SNUTP 07/015}
\end{flushright}

\title{Galactic $511$ keV line from MeV millicharged dark matter}

\author{Ji-Haeng Huh, Jihn E. Kim\footnote{jekim@phyp.snu.ac.kr},
Jong-Chul Park\footnote{jcpark@phya.snu.ac.kr} and Seong Chan Park\footnote{spark@phya.snu.ac.kr}}
\address{ Department of Physics and Astronomy, Seoul National University, Seoul 151-747, Korea
}

\begin{abstract}
We present a possible explanation of the recently observed 511 keV
$\gamma$-ray anomaly with a new ``millicharged'' fermion. The new
fermion is light (${\cal O}({\rm MeV})$) but has never been
observed by any collider experiments mainly because of its tiny
electromagnetic charge $\varepsilon e$. We show that constraints
from its relic density in the Universe and collider experiments
allow a parameter range such that the $511$ keV cosmic
$\gamma$-ray emission from the galactic bulge may be due to
positron production from this millicharged fermion.
\end{abstract}

 \pacs{95.35.+d, 95.85.Ry, 98.70.Rz, 12.60.Cn}
  \keywords{Milli-charged particles, 511 keV line, Dark matter, Galactic center}

\maketitle

The SPI/INTEGRAL observation of the very sharp $\gamma$-ray peak
at 511 keV from the galactic
bulge~\cite{Knodlseder:2003sv,spectral} needs an explanation of
its origin. Most probably, it may come from the positronium decay.
For this explanation of the positronium decay, a sufficient number
of positrons are needed in the first place. The positron abundance
in the galaxy can arise from several origins.

Some obvious candidates are the astrophysical production
mechanisms of positrons discussed in~\cite{Dermer}. However, these
mechanisms through astrophysical sources such as black holes and
supernovae turn out to be inappropriate to explain the intensity
of the positron annihilation flux, especially in emission region,
because the astrophysical sources like black holes and supernovae
are expected to be more spread out than observed. Therefore, now
most preferred interpretations of the 511 keV $\gamma$-rays rely
on particle physics origins where new particles beyond the
standard model (SM) are introduced.\footnote{We have noticed a
recent claim that the 511 keV line distribution reported in the
newest result from INTEGRAL seems to resemble the lopsided
distribution of the ``hard'' low mass X-ray binaries (LMXBs) (low
mass x-ray binaries with strong emission at $E_\gamma > 20$
keV)~\cite{Weidenspointner:2008zz}. However, quantitatively
improved understanding of 511 keV gamma ray flux coming from LMXBs
is required to see if the LMXB can fully account the anomaly. More
observation would also be required for this issue.} Usually a new
particle in the mass range $1-100$ MeV is
introduced~\cite{Boehm:2003ha}.\footnote{See also
~\cite{Finkbeiner} where ${\cal O}(100)$ GeV weakly interacting
massive particles are considered.} Let us call this new particle
$\chi$. Recent analysis including the internal bremsstrahlung
radiation and in-flight annihilation gives more stringent mass
bound for the light particle in MeV region: $m \lesssim 3-4$
MeV.\footnote{This constraint can be released by a factor of two
by a possible ionization of the propagation medium~\cite{lowerm}.}
The needed positron abundance may arise from the $\chi$ decay
and/or $\chi-\bar\chi$ annihilation to $e^+e^-$. The new light
particle should have negligible couplings to photon and $Z$ boson;
otherwise, it must have been observed at the LEP experiments. If
the new particle is neutral under the gauge transformations of the
SM as a heavy neutrino, it overcloses the Universe as noted by Lee
and Weinberg~\cite{LeeWein}. Thus, we exclude the neutrino
possibility toward the origin of the 511 keV line. This has led to
a new particle, coupling to another gauge boson beyond the SM,
e.g. as in Ref.~\cite{Fayet:2004bw}.

If another light $U(1)$ gauge boson, which will be called {\it
``exphoton,''}\footnote{In the literature, the term ``paraphoton''
is commonly used. However, we use ``exphoton'' to emphasize the
word ``extra'' which only directly couples to the ``extra'' matter
field $\chi$ and it is the gauge boson of the ``extra'' $U(1)$.
Moreover, this can show the fact that the extra $E'_8$ gauge group
may contain exphoton in heterotic string models.} beyond the SM
exists, most probably a kinetic mixing can exist via loop
effects~\cite{Okun:1982xi} between photon and exphoton without
violating the charge conservation principle. After a proper
diagonalization procedure of the kinetic energy terms, then the
electromagnetic charge of $\chi$ can be millicharged. In heterotic
string models, the extra $E_8'$ gauge group may contain the
exphoton, leading to the kinetic mixing~\cite{Dienes}. Indeed, an
explicit model for this kind from string exists in the
literature~\cite{KimJE:2007}.

Very light (${\cal O}$(eV)) millicharged particles with a
sufficiently small charge are phenomenologically acceptable as
studied in recent papers~\cite{milliphen}. On the other hand, the
heavy millicharged particle idea as a dark matter (DM) candidate
was suggested about 20 years ago~\cite{Goldberg:1986nk} and it has
been revived recently~\cite{Cheung:2007ut}. The intermediate
${\cal O}$(MeV) millicharged particles has not been ruled out by
observations in the previous study~\cite{Davidson:2000hf} which,
however, did not include the $511$ keV line possibility. Earlier,
the ${\cal O}$(MeV) millicharged particle effect on cosmic
microwave background radiation was studied in the parameter region
of exphoton coupling constant ($\alpha_{\rm ex} \equiv e^2_{\rm
ex}/4\pi \sim 0.1$)~\cite{Dubovsky}. Here, we analyze the urgent
problem of the ${\cal O}$(MeV) millicharged particles toward
interpreting the 511 keV line within the limit provided by the DM
constraint with reasonable exphoton coupling
constants.\footnote{The laboratory and cosmological bound of
millicharged particles was studied sometime
ago~\cite{Davidson:2000hf}, but the study toward 511 keV line and
the subeV mass range has not been included.}

Consider two Abelian gauge groups $U(1)_{\rm QED}$ and $U(1)_{\rm
ex}$.\footnote{One should note that the $U(1)$ mixing in the
observable and hidden sectors should be considered carefully. For
the simple assumption of the charges given in
Ref.~\cite{Ahlers:2007rd}, $\chi$ coupling with the full strength
to the massive exphoton does not couple to the massless photon, or
at least suppressed by $\varepsilon$. Converting this argument,
the massive $Z$-boson mixing with the massless exphoton gives the
neutrino coupling to the exphoton suppressed by $\varepsilon$.
Thus, the very stringent supernova cooling constraint which gives
a bound for the low-energy dark matter ($m > 10$
MeV)~\cite{Fayet:2006sa} does not apply to our case since
$\nu\chi$ cross section is suppressed by $\varepsilon^2$ compared
to that of~\cite{Fayet:2006sa}.} The kinetic mixing of $U(1)_{\rm
QED}$ photon and $U(1)_{\rm ex}$ exphoton is parameterized as
\begin{eqnarray}
{\cal L} = -\frac{1}{4}\hat{F}_{\mu\nu}\hat{F}^{\mu\nu}
-\frac{1}{4}\hat{X}_{\mu\nu}\hat{X}^{\mu\nu}
-\frac{\xi}{2}\hat{F}_{\mu\nu}\hat{X}^{\mu\nu}, \label{lagrangian}
\end{eqnarray}
where $\hat{A}_\mu (\hat{X}_{\mu})$ is the $U(1)_{\rm QED}
(U(1)_{\rm ex})$ gauge boson and its field strength tensor is
$\hat{F}_{\mu\nu} (\hat{X}_{\mu\nu})$. The kinetic mixing is
parameterized by $\xi$ which is generically allowed by the gauge
invariance and the Lorentz symmetry. In the low-energy effective
theory, $\xi$ is considered to be a completely arbitrary
parameter. An ultraviolet theory is expected to generate the
kinetic mixing parameter $\xi$~\cite{Okun:1982xi}. The usual
diagonalization procedure of these kinetic terms leads to the
relation,
\begin{eqnarray}
\left(
  \begin{array}{c}
    A_\mu \\
    X_\mu \\
  \end{array}
\right) = \left(
            \begin{array}{cc}
              \sqrt{1-\xi^2} & 0 \\
              \xi & 1 \\
            \end{array}
          \right) \left(
                    \begin{array}{c}
                      \hat{A}_\mu \\
                      \hat{X}_\mu \\
                    \end{array}
                  \right),
\end{eqnarray}
and we obtain
\begin{align}
{\cal L} =
-\frac{1}{4}F_{\mu\nu}F^{\mu\nu}-\frac{1}{4}X_{\mu\nu}X^{\mu\nu},
\end{align}
where the new field strengths are  $F_{\mu\nu}$ and $X_{\mu\nu}$.
Photon corresponds to $A_\mu$ and {\it exphoton} corresponds to
$X_\mu$. If the exphoton is exactly massless, there exists an
$SO(2)$ symmetry in the $A_\mu-X_\mu$ field space:
$A_\mu\to\cos\theta A_\mu+\sin\theta X_\mu$ and
$X_\mu\to-\sin\theta A_\mu+\cos\theta X_\mu$. Any physical
observable, however, does not depend on $\theta$.

Using the above $SO(2)$ symmetry, let us take the following simple
interaction Lagrangian of a SM fermion, i.e. electron, with a
photon in the original basis as
\begin{align}
{\cal L}= \bar{\psi} \left( \hat{e} Q \gamma^\mu \right) \psi
\hat{A}_\mu. \label{SM-interaction}
\end{align}
Note that in this basis there is no direct interaction between the
electron and the hidden sector gauge boson $\hat{X}$. If there
exists a hidden sector Dirac fermion $\chi$ with the $U(1)_{\rm
ex}$ charge $Q_{\chi}$, its interaction with the hidden sector
gauge boson is simply represented by
\begin{align}
{\cal L}= \bar{\chi} \left( \hat{e}_{\rm ex} Q_{\chi} \gamma^\mu
\right) \chi \hat{X}_{\mu}, \label{hidden-interaction}
\end{align}
where $\hat{e}_{\rm ex}$ can be different from $\hat{e}$ in
general. In this case, there is also no direct interaction between
the hidden fermion and the visible sector gauge boson $\hat{A}$.
We can recast the Lagrangian~(\ref{SM-interaction}) in the
transformed basis $A$ and $X$,
\begin{align}
{\cal L}= \bar{\psi} \left( \frac{\hat{e}}{\sqrt{1-\xi^2}} Q \gamma^\mu \right) \psi A_\mu.
\end{align}
Here, one notices that the standard model fermion has a coupling
only to the visible sector gauge boson $A$ even after changing the
basis of the gauge bosons. However, the coupling constant
$\hat{e}$ is modified to $\hat{e} / \sqrt{1-\xi^2}$, and so the
physical visible sector coupling $e$ is defined as $e \equiv
\hat{e} / \sqrt{1-\xi^2}$. Similarly, we derive the following for
$\chi$,
\begin{align}
{\cal L} = \bar{\chi} \gamma^\mu \left(\hat{e}_{\rm ex} Q_{\chi} X_\mu
-\hat{e}_{\rm ex}\frac{\xi}{\sqrt{1-\xi^2}} Q_{\chi} A_\mu \right) \chi .
\label{hidden-shift}
\end{align}
In this basis, the hidden sector matter field $\chi$ now can
couple to the visible sector gauge boson $A$ with the coupling
$-\hat{e}_{\rm ex} \xi / \sqrt{1-\xi^2}$. In terms of the
aforementioned $SO(2)$ symmetry, it simply means the mismatch
between the gauge couplings of the electron and other fermions.
Thus, we can set the physical hidden sector coupling $e_{\rm ex}$
as $e_{\rm ex} \equiv \hat{e}_{\rm ex}$ and we define the coupling
of the field $\chi$ to the visible sector gauge boson $A$,
introducing the millicharge parameter $\varepsilon$, as
$\varepsilon e \equiv -{e}_{\rm ex} \xi / \sqrt{1-\xi^2}$. Note in
general that $e\ne e_{\rm ex}$. Since $\xi \simeq \varepsilon
e/e_{\rm ex}$ is expected to be small, the condition $\xi<1$ gives
$\alpha_{\rm ex}/{\alpha} > \varepsilon^2$. From a fundamental
theory, one can calculate the ratio $e_{\rm ex}/e$ in principle,
which is possible with the detail knowledge of the
compactification radius~\cite{KKmasses}. Here, we simply take the
ratio as a free parameter.

\begin{figure}[t]
\begin{center}
\includegraphics[width=0.80\linewidth]{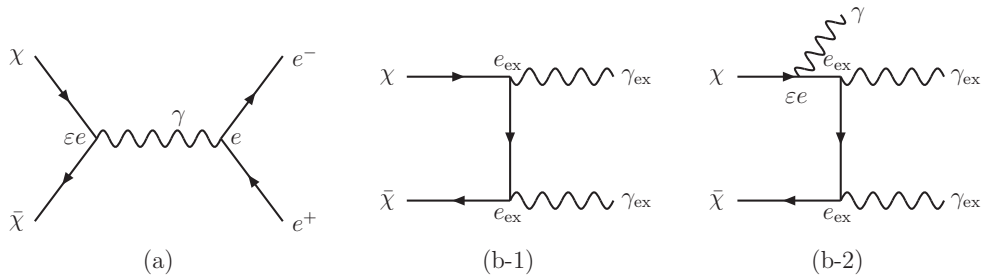}
\end{center}
\caption{The ``millicharge'' annihilation diagrams to, (a)
$e^+e^-$ and (b-1) $2\gamma_{\rm ex}$; (b-2) the bremsstrahlung
diagram related to (b-1). The cross diagram in (b-1) is not
shown.}\label{feynman}
\end{figure}

For the cosmological study of $\chi$, we need the annihilation
cross sections of DM: $\chi\bar{\chi}\rightarrow e^-e^+$,
$\chi\bar{\chi}\rightarrow 2\gamma_{\rm ex}$,
$\chi\bar{\chi}\rightarrow \gamma\gamma_{\rm ex}$, and
$\chi\bar{\chi}\rightarrow \gamma \gamma$. The ratio for these
cross sections is given by
\begin{eqnarray} \sigma_{2\gamma_{\rm
ex}}:\sigma_{e+e-}:\sigma_{\gamma \gamma_{\rm
ex}}:\sigma_{2\gamma} \simeq  \alpha_{\rm ex}^2: \varepsilon^2
\alpha^2:\varepsilon^2 \alpha\alpha_{\rm ex}:\varepsilon^4
\alpha^2.
\end{eqnarray}
We noticed that the first two channels (depicted in
Fig.~\ref{feynman}) are important and the last two channels are
quite suppressed in the parameter region where $\varepsilon$ and
$\alpha_{\rm ex}/\alpha$ are quite {\it small} as is required by
the observational data. If $\alpha_{\rm ex}/\alpha > 0.01 (0.1)$,
the background diffuse gamma-ray flux could be larger than $1
(10)\%$ of the 511 keV flux, so the region is already excluded by
the INTEGRAL and COMPTEL measurements~\cite{lowerm, COMPTEL} (See
Fig.~\ref{relic}). As we will see below,
$\chi\bar{\chi}\rightarrow 2\gamma_{\rm ex}$ channel ((b-1) in
Fig.~\ref{feynman}) overwhelmingly dominates in the first two main
channels of Fig.~\ref{feynman}. Then it seems that the gamma-ray
flux from the real bremsstrahlung ((b-2) in Fig.~\ref{feynman})
could be of considerable amount. However, the bremsstrahlung cross
section is suppressed by a factor of $\varepsilon^2\alpha$
compared to that of diagram (b-1) in Fig.~\ref{feynman}. Thus,
$\sigma^{\rm brem}_{2\gamma_{\rm ex}} \sim \alpha (\alpha_{\rm
ex}/\alpha)^2 \sigma_{e^+e^-}$ and is negligible. The annihilation
cross sections determine the relic density of the hidden sector
fermion $\chi$. The process $\chi\bar{\chi}\rightarrow e^-e^+$
determines the flux of the eventual $511$ keV photons as well. Let
us assume that the charge of the $\chi$ particle is  $(0,
\hat{e}_{\rm ex})$ in the basis of $(\hat{A}, \hat{X})$. The
millicharge $\varepsilon e$ comes from the shift of the exphoton
field in Eq.~(\ref{hidden-shift}) and $e_{\rm ex}$ is for the
hidden sector $U(1)_{\rm ex}$ gauge interaction.

The cross section for the process $\chi\bar\chi\to e^- e^+$, shown
in Fig.~\ref{feynman}(a), is given by
\begin{align}
\sigma_{\chi\bar{\chi}\rightarrow
e^-e^+}=\frac{4\pi}{3}\frac{\varepsilon^2
\alpha^2}{s}\frac{\beta_e}{\beta_{\chi}}
\left[1+2\frac{m_e^2+m_{\chi}^2}{s}
+\frac{4m_e^2m_{\chi}^2}{s^2}\right],
\end{align}
where $\beta_i = \sqrt{1-4 m_i^2/s}$ is the velocity of the
particle-$i$ and $\alpha \equiv e^2/4\pi$. In the nonrelativistic
regime, the approximation $E\sim
m_{\chi}+\frac{1}{2}m_{\chi}(v_{\textrm{rel}}/2)^2$ makes sense
and we obtain
\begin{align}
\sigma_{\chi\bar{\chi}\rightarrow e^-e^+}
=\pi\varepsilon^2\alpha^2\frac{1}{m_{\chi}^2}
\frac{1}{v_{\textrm{rel}}}\left[1-\frac{m_e^2}{m_{\chi}^2}
\right]^{1/2}\left[1+\frac{m_e^2}{2m_{\chi}^2}\right]
+\cdots .\label{sigma-ee}
\end{align}
Now, the cosmologically interesting average of the cross section
times velocity, $\langle\sigma v\rangle_{e^-e^+}$, becomes
$\langle\sigma v\rangle_{e^-e^+} = a_{e^-e^+}+b_{e^-e^+} \langle
v^2\rangle + {\cal O}(\langle v^4 \rangle )$ where $a_{e^-e^+}$
and $b_{e^-e^+}$ are given by Eq.~(\ref{sigma-ee}),
\begin{align}
a_{e^-e^+} &= \frac{\pi \varepsilon^2
\alpha^2}{m_{\chi}^2}\left[1-
\frac{m_e^2}{m_{\chi}^2}\right]^{1/2}
\left[1+\frac{1}{2}\frac{m_e^2}{m_{\chi}^2}\right],
\nonumber \\
b_{e^-e^+} &= \frac{23\pi \varepsilon^2
\alpha^2}{96m_{\chi}^2}\left[1-\frac{m_e^2}{m_{\chi}^2}
\right]^{-1/2}\left[\frac{59}{46}
\frac{m_e^4}{m_{\chi}^4}+\frac{1}{2}
\frac{m_e^2}{m_{\chi}^2}-1\right].
\end{align}

Similarly, for the process $\chi\bar{\chi} \rightarrow
2\gamma_{\rm ex}$ shown in Fig.~\ref{feynman}(b-1), we obtain
\begin{align}
\frac{d\sigma_{\chi\bar{\chi} \rightarrow 2\gamma_{\rm ex}}}{d{\rm
cos}\theta} =\frac{2\pi\alpha_{\rm
ex}^2}{s\beta_{\chi}}\left[\frac{1+2\beta^2_{\chi}{\rm
sin}^2\theta-\beta^4_{\chi}(2{\rm sin}^2\theta+{\rm
cos}^4\theta)}{(1-\beta^2_{\chi}{\rm cos}^2\theta)^2}\right],
\label{sigma-rr}
\end{align}
where $\alpha_{\rm ex} \equiv e^2_{\rm ex}/4\pi$. The total cross
section is given by $\sigma_{\chi\bar{\chi} \rightarrow
2\gamma_{\rm ex}}=\int^1_0 d({\rm cos} \theta)
\frac{d\sigma}{d{\rm cos}\theta}$. In this case also, the
cosmological average of the annihilation cross section times
velocity, $\langle\sigma v\rangle_{2\gamma_{\rm ex}}$, is
expressed in powers of $v^2$ as $\langle\sigma
v\rangle_{2\gamma_{\rm ex}} = a_{2\gamma_{\rm ex}}+b_{2\gamma_{\rm
ex}} \langle v^2 \rangle + {\cal O}(\langle v^4 \rangle )$ where
$a_{2\gamma_{\rm ex}} = {\pi \alpha^2_{\rm ex}}/{m_{\chi}^2}$ and
$b_{2\gamma_{\rm ex}} = \frac{11}{32}a_{2\gamma_{\rm ex}}.$ Again,
we neglected the contributions from $\chi\bar{\chi}\rightarrow
\gamma\gamma_{\rm ex}, \gamma\gamma$ because of the smallness of
$\varepsilon$ and $\alpha_{\rm ex}/\alpha$.

The relic density of a generic relic, $X$, can be expressed as
\begin{equation}
\begin{split}
\Omega_X h^2 & \approx \frac{1.07 \times 10^9 \, {\rm GeV}^{-1}}
{M_{Pl}}\frac{x_F}{\sqrt{g_*}}\frac{1}{(a+3b/x_F)}\\
& \approx 8.77 \times 10^{-17} \, {\rm
MeV}^{-2}\frac{x_F}{\sqrt{g_*}}\frac{1}{(a+3b/x_F)},
\end{split}
\end{equation}
where $g_*$ is evaluated at the freeze-out temperature $T_F$, $a$
and $b$ are the velocity independent and dependent coefficients,
respectively, and $x_F=m_X/T_F \simeq17.2 +
\ln(g/g_*)+\ln(m_X/{\rm GeV})+\ln\sqrt{x_F} \sim 12-19$ for
particles in the MeV--GeV range~\cite{Bertone:2004pz}. We can
approximate $x_F \approx 11.6+\ln(m_X/{\rm MeV})$ for 1 MeV
$\lesssim m_X \lesssim 100$ MeV. Therefore, we can estimate the
relic density of the millicharged particle, $\chi$, as
\begin{equation}
\Omega_{\chi} h^2 \approx 1.60 \times 10^{-13} \, \frac{(11.6+\ln
\overline{m}) \overline{m}^2}{\left(\frac{\alpha_{\rm ex}}{\alpha}
\right)^2+\varepsilon^2\left(1-\frac{m^2_e}{m^2_{\chi}}
\right)^{1/2}\left(1+\frac{m^2_e}{2m^2_{\chi}}\right)},
\end{equation}
where $\overline{m}\equiv m_{\chi}/{\rm MeV}$ and we put
$g_*\simeq 10.75$ for $1<T_F/{\rm MeV}<100$.\footnote{In this
step, we use the total annihilation cross section, i.e.
$a=a_{e^-e^+}+a_{2\gamma_{\rm ex}}$ and
$b=b_{e^-e^+}+b_{2\gamma_{\rm ex}}$.} Finally, we can find a
constraint for the mass $m_{\chi}$ and the charge $\varepsilon$ of
the millicharged DM and the hidden sector coupling $\alpha_{\rm
ex}$, based on the relic density of DM from the WMAP three-year
results~\cite{Spergel:2006hy}. In Fig.~\ref{relic}, we present the
excluded parameter space for typical DM masses ($m_\chi = 1, 3$
and $10$ MeV) as the yellow shaded regions from our analysis of
the DM relic density.

\begin{figure}[t]
\begin{center}
\includegraphics[width=0.65\linewidth]{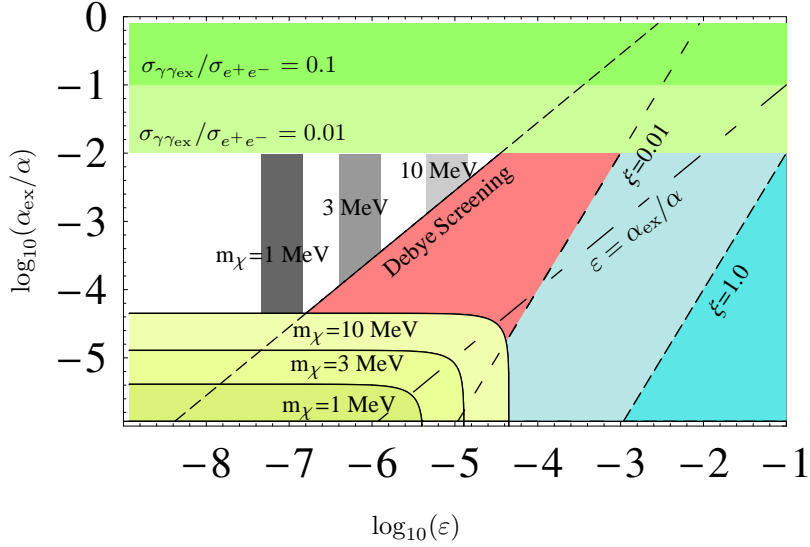}
\end{center}\caption{
The plot for $\alpha_{\rm ex}/\alpha$ versus $\varepsilon$. The
lower left corner (yellow shaded regions) is excluded by the DM
relic density constraint: the lines correspond to
$\Omega_{\chi}h^2=0.11$ and $m_{\chi}=1, 3,$ and $10$ MeV,
respectively. The vertical bands (gray shaded regions) are the
allowed range of $\varepsilon$ that will be given by the 511 keV
$\gamma$-ray flux constraint analysis: the regions correspond to
$m_{\chi}=1, 3,$ and $10$ MeV, respectively. The region excluded
by the Debye screening is shown from the central (pink shaded)
region to the lower right corner marked by Debye screening. The
(green) region $\alpha_{\rm ex}/\alpha
>0.01 (0.1)$ is excluded since more than $1 (10)\%$ diffuse gamma
ray flux compared to the 511 keV flux is expected.} \label{relic}
\end{figure}

In charged medium, photon can effectively obtain mass via the
interaction with charged particles. Therefore, this effective mass
should be smaller than the limit of photon mass. As a result, the
Debye screening length in the DM medium around Earth $\lambda_D=
\sqrt{T_\chi/ \varepsilon^2 e^2 n_\chi}$ is required to be larger
than the limit of the inverse photon
mass~\cite{Williams:1971ms,Mitra:2006ds}. Putting $n_\chi =
\rho_\chi/m_\chi \simeq 0.3 {\rm GeV/cm^3} \times
\Omega_\chi/(\Omega_{\rm DM}m_\chi)$ and $\Omega_{\rm DM}\simeq
0.23$, we finally get the simple relation $\frac{\alpha_{\rm
ex}}{\alpha} \gtrsim 282 \varepsilon$. One should note that the
relic density of $\chi$ is essentially proportional to $m_\chi^2$
so that the Debye screening length is not sensitive to the mass.
The lower right corner from the central region (colored by pink)
is excluded by this constraint. Interestingly, $m_\chi \gtrsim 3$
MeV does not have the parameter space which can fully accommodate
the dark matter density $\Omega_{\rm DM}\simeq 0.23$.

The line $\varepsilon = \alpha_{\rm ex}/\alpha$ corresponds to the
line of equal couplings that divides where the diagrams (a) and
(b-1) in Fig.~\ref{feynman} dominate: in the upper part of the
line the process $\chi\bar{\chi}\rightarrow 2\gamma_{\rm ex}$ and
in the lower part the process $\chi\bar{\chi}\rightarrow e^-e^+$
dominate. In addition, we show the allowed range of $\varepsilon$
for typical DM masses ($m_\chi = 1, 3$ and $10$ MeV) as the (gray
shaded) vertical bands, which will be obtained from the following
analysis of the 511 keV $\gamma$-ray flux constraint. For example,
if $m_\chi=3$ MeV, the middle (gray shaded) vertical band for
$\varepsilon$ in the upper left corner is allowed. The smallness
requirement of $\xi$ is buried in the Debye screening length
constraint. The study of~\cite{Dubovsky} is buried in the lower
right corner around $\xi=1$. As can be seen from the figure, a
significant region is excluded. However, we note that there still
remains an available space.

The observed flux of dark matter annihilation products can be
obtained by integrating the density squared along the line of
sight as
\begin{equation}
\Phi_i(\psi, E)=\sigma v \frac{dN_i}{dE} \frac{1}{4 \pi
m_{\rm{DM}}^2} \int_{\mbox{line of sight}}d s
\rho^2\left(r(s,\psi)\right) , \label{phi-i}
\end{equation}
where $\rho(r)$ is the mass density of the DM, $\sigma$ is the DM
annihilation cross section, $v$ is the velocity, $dN_i/dE$ is the
spectrum of secondary particles of species $i$, and $s$ is the
coordinate running along the line of sight, in a direction making
an angle, $\psi$, from the direction of the galactic center. It is
convenient to introduce the quantity $J(\psi)$~\cite{Bergstrom98}:
\begin{equation} J\left(\psi\right) = \frac{1}
{8.5\, \rm{kpc}} \left(\frac{1}{0.3\,
\mbox{\small{GeV/cm}}^3}\right)^2 \int_{\mbox{\small{line of
sight}}}d s \rho^2\left(r(s,\psi)\right)\,
\end{equation}
by which the expression in eq.~(\ref{phi-i}) can be separated to
``halo profile depending'' factors and ``particle physics
depending'' factors as
\begin{equation} \Phi_{i}(\Delta\Omega, E)\simeq5.6
\,\frac{dN_i}{dE} \left( \frac{\sigma v}
{\rm{pb}}\right)\left(\frac{1\rm{MeV}} {m_{\rm{DM}}} \right)^2
\overline{J}\left(\Delta\Omega\right)
 \; \Delta\Omega\,\,\rm{cm}^{-2} \rm{s}^{-1}\,
\end{equation}
where $\overline{J}(\Delta\Omega)$ is defined as  the average of
$J(\psi)$ over a spherical region of solid angle, $\Delta\Omega$,
centered on $\psi=0$~\cite{Bertone:2004pz}.

If the mass of the DM particle is less than the muon mass, the low
velocity annihilations can produce electron-positron pairs. Most
positrons lose energy through their interactions with the inter
stellar medium (ISM) and bremsstrahlung radiation and go rest.
Then positron annihilation takes place via the positronium
formation ($\sim 96.7\pm 2.2 \%$)~\cite{spectral} and partly via
the direct annihilation into two $511$ keV gamma-rays. Only $25\%$
of the time, a singlet positronium state decaying to two $511$ keV
photons is formed while $75\%$ of the time, a triplet state
decaying to three continuum photons is formed. This means that the
$511$ keV photon emission occurs only by a quarter of the total
positron production through DM annihilation. After taking all this
into account, the flux of $511$ keV $\gamma$-rays from the
galactic center can be given as
\begin{equation}
\Phi_{\gamma,511} \simeq 0.275 \times 5.6 \, \bigg(\frac{\sigma v}
{{\rm pb}}\bigg) \bigg(\frac{1\,\rm{MeV}}{m_{\chi}}\bigg)^2
\overline{J}(\Delta \Omega) \Delta \Omega \, {\rm cm}^{-2} {\rm
s}^{-1}, \label{flux}
\end{equation}
where $\Delta \Omega$ is the observed solid angle toward the
direction of the galactic center.

The observed $\gamma$-ray profile has a full width at half maximum
of $\sim6^{\circ}$ with a $4^{\circ}-9^{\circ}$ 2$\sigma$
confidence interval and the flux $\Phi_{\gamma,511} \simeq (1.02
\pm 0.10) \times 10^{-3}$ ph cm$^{-2}$
s$^{-1}$~\cite{Knodlseder:2003sv,spectral}. Thus, we consider a
solid angle of $0.0086$ sr, corresponding to a $6^{\circ}$
diameter circle. In this model, positron is produced from the
process $\chi\bar{\chi}\rightarrow e^-e^+$. Therefore, we can find
the charge $\varepsilon$ of the millicharged DM as a function of
its mass $m_{\chi}$ from the resultant cross section
$\langle\sigma v\rangle_{e^-e^+}$ for this process and
Eq.~(\ref{flux}). The relation is given by
\begin{eqnarray}
\varepsilon \simeq 1.0
\times10^{-6}\frac{\overline{m}^2}{\sqrt{\overline{J}}}
\left[1-\frac{m^2_e}{m^2_{\chi}}\right]^{-1/4}
\left[1+\frac{m^2_e}{2m^2_{\chi}}\right]^{-1/2},
\end{eqnarray}
where $\overline{m}\equiv m_{\chi}/{\rm MeV}$. To estimate the
required parameter space, we use the width of the observed
distribution $\overline{J}(0.0086{\rm\ sr}) \sim 50-500$,
approximately corresponding to $\gamma \simeq 0.6-1.2$ essentially
following the approach of Ref.~\cite{Boehm:2003bt}.\footnote{If
the main source of 511 keV $\gamma$-rays from galactic bulge is
from the DM  annihilation, the observed distribution of 511 keV
emission line would constrain the shape of the DM halo profile
because DM annihilation rate is proportional to the DM density
squared.}

There already exist various bounds from experimental and
observational results, which are summarized
in~\cite{Davidson:2000hf}. Among them, the limit from the
millicharged particle search experiment at
SLAC~\cite{Prinz:1998ua} is relevant to the mass-charge parameter
space, which is considered in this analysis. In principle, the DM
can contribute to the anomalous magnetic moment~\cite{g-2}, but it
can only occur at the two-loop level with an additional
$\varepsilon^2$ suppression factor. The expected recoil energy by
the DM-nucleon scattering is too small to be measured by the
existing or near-future experiments because of the lightness of
the proposed DM candidate.

\begin{figure}[t]
\begin{center}
\includegraphics[width=0.65\linewidth]{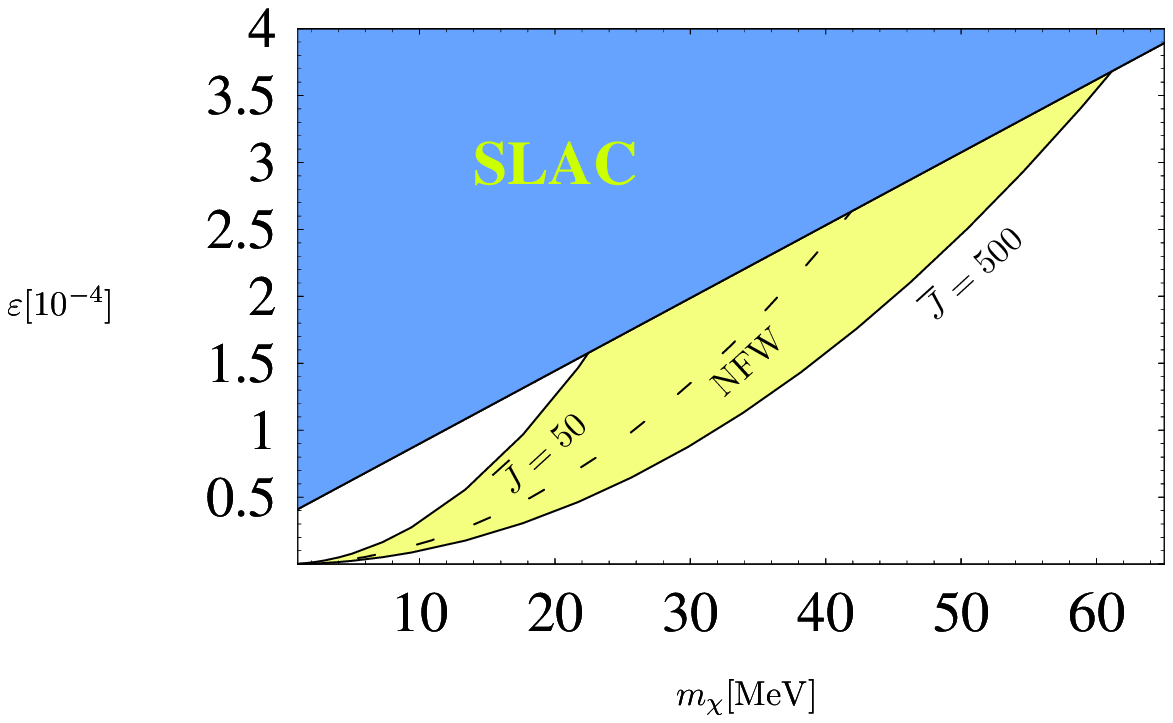}\\
\vskip 0.2cm
\includegraphics[width=0.65\linewidth]{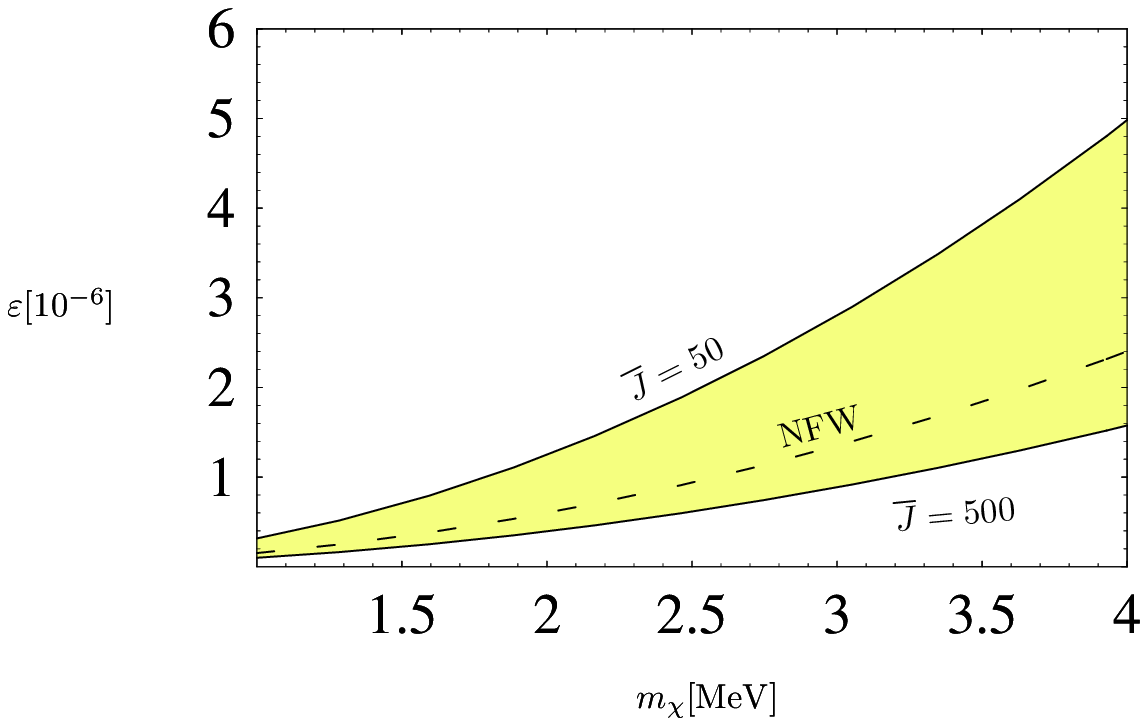}
\end{center}\caption{The plot for $\varepsilon$ versus $m_{\chi}$.
The dark (blue) shaded region is excluded by the SLAC search of
millicharged particles~\cite{Prinz:1998ua}. We plot the
Navarro-Frenk-White profile~\cite{Navarro} model line (dotted)
using the fitting parameter ($R=20$ kpc, $\rho_0 = 0.347$
GeV/${\rm cm}^3$) and the lightly shaded (yellow) region for the
uncertainty range $\overline{J}(0.0086{\rm\ sr}) \sim 50-500$.
After considering the recently given strong constraint on the
light dark matter mass~\cite{lowerm}, the allowed region is
$m_\chi \lesssim 4$ MeV (bottom). } \label{511flux}
\end{figure}

The result from the study of the 511 keV $\gamma$-ray flux and the
SLAC experiment is presented in Fig.~\ref{511flux} in the
$\varepsilon-m_\chi$ space. Even after taking into account the
SLAC bound for the millicharged particle, a large parameter region
is still remaining. Recent analysis such as the internal
bremsstrahlung radiation and in-flight annihilation gives strong
mass bound for the light dark matter in MeV region: $m \lesssim
3-4$ MeV.\footnote{As already stated in the beginning, this
constraint can be reduced by a factor of two by a possible
ionization of the medium~\cite{lowerm}.} Therefore, the lower left
corner is magnified. In the allowed parameter region ($1 \gg
\alpha_{\rm ex}/\alpha > \varepsilon$), the relic density of DM is
essentially determined by $\chi\bar{\chi}\rightarrow 2\gamma_{\rm
ex}$. However, the observed $511$ keV photon flux is mostly
explained by $\chi\bar{\chi}\rightarrow e^-e^+$. In this respect,
the difficulty of explaining both quantities in
Ref.~\cite{Boehm:2003bt} is easily avoided in our model.

One final comment is about the spontaneously broken $U(1)_{\rm
ex}$ symmetry which results in the nonvanishing exphoton mass. In
this case, the electrically charged particles such as electron and
proton can couple to the exphoton though the hidden fermion
($\chi$) does not directly couple to the on-shell
photon~\cite{Ahlers:2007rd}. In principle, this case can be also
relevant to our DM problem and the related $511$ keV photon line.
Theoretically, spontaneous symmetry breaking generally gives
finite ranges of parameter space both for massless and massive
exphotons and hence our study on massless exphoton covers a finite
range of the parameter space. In the future, we would like to
discuss the cosmology of ${\cal O}$(MeV) exphoton.

In conclusion, we presented an allowed parameter range of a new
millicharged particle $\chi$ with ${\cal O}$(MeV) mass toward a
possible solution to the recently observed $511$ keV cosmic
$\gamma$-ray anomaly. It couples to photon with a ``milli''
electric charge strength, $\varepsilon e$. In the mass range of
$m_\chi \lesssim 1$ MeV, the millicharged particle can constitute
a sizable ($\gtrsim 10\%$) portion of the DM content of the
Universe but might have escaped detection so far in any collider
experiments basically because of its tiny electric charge. This
millicharged particle may arise in a more fundamental theory such
as string as an interplay between the observable and hidden
sectors.

\acknowledgements{This work was supported in part by the Korea Research Foundation Grant funded by
the Korean Goverment(MOEHRD) (KRF-2005-084-C00001 and No.R14-2003-012-01001-0). S.C.P. is
supported by the BK21 program of Ministry of Education.}

\end{document}